# An Efficient Method for Image and Audio Steganography using Least Significant Bit (LSB) Substitution

Ankit Chadha, Neha Satam, Rakshak Sood, Dattatray Bade
Electronics and Telecommunication Department,
Vidyalankar Institute of Technology,
Mumbai, India.

## ABSTRACT
In order to improve the data hiding in all types of multimedia data formats such as image and audio and to make hidden message imperceptible, a novel method for steganography is introduced in this paper. It is based on Least Significant Bit (LSB) manipulation and inclusion of redundant noise as secret key in the message. This method is applied to data hiding in images. For data hiding in audio, Discrete Cosine Transform (DCT) and Discrete Wavelet Transform (DWT) both are used. All the results displayed prove to be time-efficient and effective. Also the algorithm is tested for various numbers of bits. For those values of bits, Mean Square Error (MSE) and Peak-Signal-to-Noise-Ratio (PSNR) are calculated and plotted. Experimental results show that the stego-image is visually indistinguishable from the original cover-image when n<=4, because of better PSNR which is achieved by this technique. The final results obtained after steganography process does not reveal presence of any hidden message, thus qualifying the criteria of imperceptible message.

**Keywords -** Steganography, data hiding, LSB manipulation, MSE, PSNR.

## 1. INTRODUCTION
Secured data communication has been a crux point since the advent of various hacking technologies. Conveying the messages is an important issue which has become a task as the number of eavesdroppers has increased. This situation led to the idea of encoding information using redundant characters and making it obscure to any third party person. Data prone to such attacks could be encoded using different methods and this gave rise to Steganography. Steganography literally means covered writing, which was practiced during ancient times. It consisted of etching messages in wooden tablets and covering them with wax so that the message was hidden and only knowledgeable person will be able to fetch the message. Though the methods in that era were physical, they served the purpose. As times evolved, more methodical practices were followed which dealt with obscuring the data itself to an extent. This marked the era of digital Steganography. Though the purpose may match with another parallel branch viz. Cryptography, there is subtle difference in them. In latter, the messages are encoded which can be clearly separated from other messages by observing the redundancies. Even if the message seems unbreakable, its encoded nature may draw suspicion and can be dangerous. As for Steganography, a seemingly innocuous image can be made to hide data which becomes imperceptible.

Watermarking is also a branch of steganography. It is “the practice of imperceptibly altering a Work to embed a message about that Work” [1]. Watermarking in images should meet the few imperative requirements as described below:

1. Imperceptibility to a Human Visual System:
   Whenever an image is subjected to watermarking process, its watermarked version should look similar to the one which is not watermarked. There should not be any palpable distinctions so that third party person might learn about watermarking. If this condition is fulfilled then it serves the primary purpose of watermarking.

2. Robust to various kinds of distortions:
   The watermarked image should be robust and sturdy against distortions such as lossy compressions or any types of modifications. It should produce the original image after reversal of processes. This makes sure that the message to be hidden remains safe even if it gets attacked and altered.

3. Simplicity of detection and extraction:
   For an individual who possesses the private key to retrieve the message, it should be easy to extract it. For others who don’t have a key, it should be very complicated to unlock its contents. This ensures that the watermarking is perfect and can be only deciphered by rightful individual.

4. High information capacity:
   The watermarked image should be able to carry large information without burdening the channel or the original image. This property describes how much data should be embedded for successful detection. The message should be well hidden and convey information in desired manner. The message hidden should not be a load onto the cover image and should not degrade its quality.

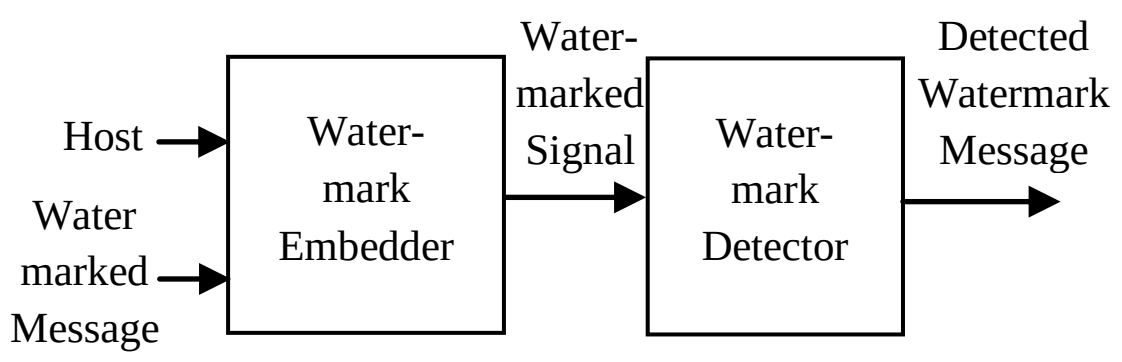

**Fig.1: General watermark block diagram**

On the similar basis, audio watermarking can be performed. There are three ways to implement audio watermarking:

1. Audio in audio:
   Both the cover and message are audio signals. Concept wise, it is similar to image in image watermarking.

2. Image in audio:
   Cover is audio and message is image signal. 1D DCT can be used for performing this type of watermarking.

3. Audio in image:
   Cover is image and message is audio signal. Similar to image in audio, 1D DCT can be used.

Figure 1 gives the overview of general watermarking system. The host signal also known as cover signal gets embedded with watermark message, a message to be hidden, by watermark embedder. It provides the secret key using which at the receiver side, the message can be decoded. Figure 2 gives illustration of image watermarking technique. This paper is organized as follows: section 2 gives idea of proposed solution and section 3 gives implementation steps. Section 4 shows experimental results and section 5 provides conclusion.

## 2. IDEA OF PROPOSED SOLUTION

The concept of modified least significant bit (LSB) substitution is to increase or decrease the most significant bit (MSB) part by 1 in order to improve the image quality [2,3]. In LSB steganography, the least significant bits of the cover image's digital data are used to mask the message. The simplest of the LSB steganography techniques is LSB substitution.

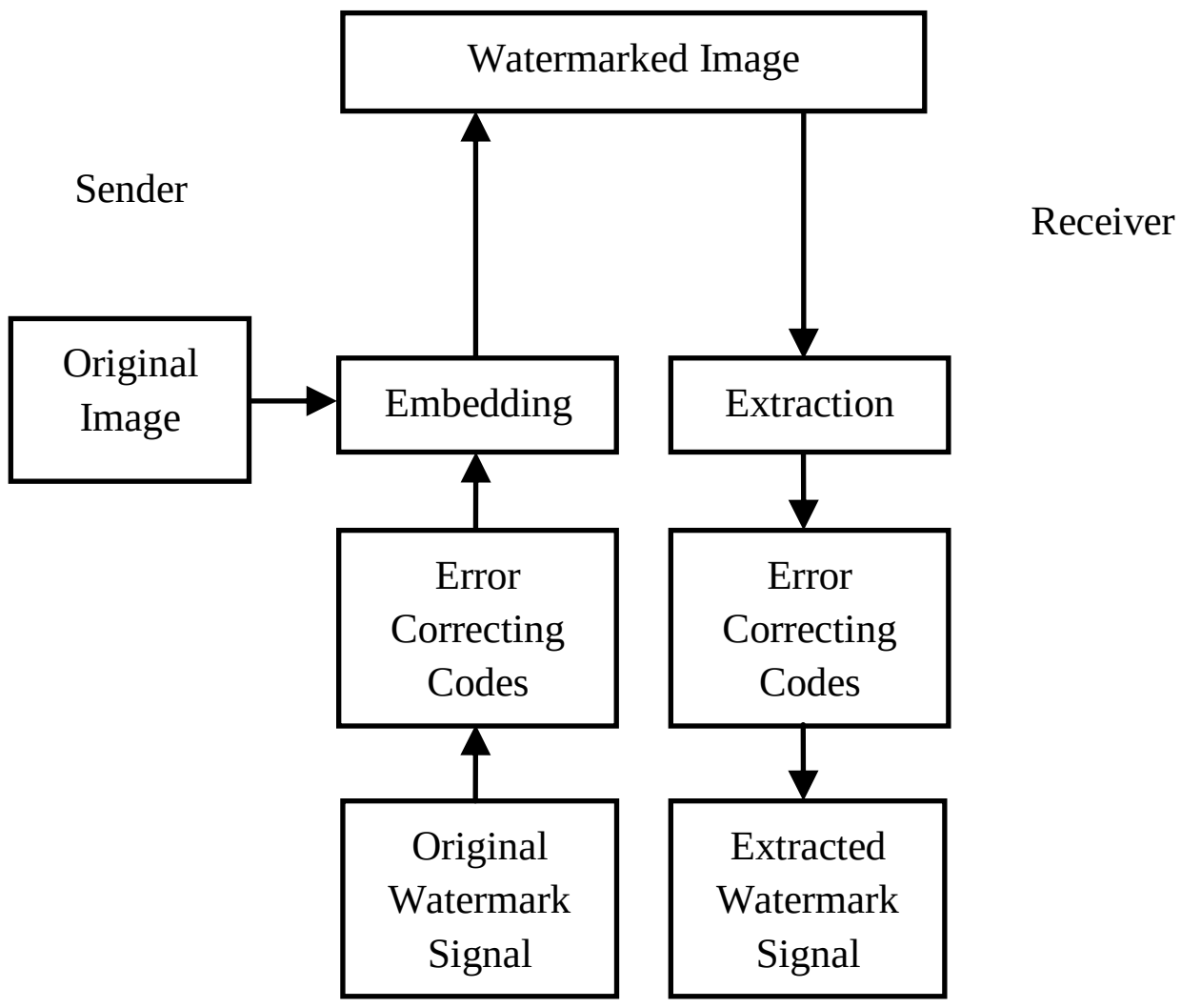

**Fig.2: Illustration of Watermarking technique**

The substitution operation $x_j[i] \rightarrow m[i]$ on the LSBs, where $x$ is cover image and $m$ is message to be hidden, is performed on this subset. The extraction process simply retrieves the watermark by reading the value of these bits. Therefore, the decoder needs all the samples of the watermarked audio that were used during the embedding process. Usually, $l(x_0) >> l(m)$, i.e., the length of cover image is larger than length of message to be hidden . Thus the robustness of the method can be improved by a repeated watermark embedding [4].

8-bit images are vulnerable to LSB manipulation because of color limitations. For that reason, the cover image must be more carefully selected so that the existence of an embedded message is virtually unknown. When information is inserted into the LSBs of the data, the pointers to the color entries in the palette are changed. Consider an 8-bit grayscale bitmap image where each pixel is stored as a byte representing a grayscale value. Suppose the first eight pixels of the original image have the following grayscale values:

10010111 10001100 11010010 01001010 00100110
01000011 00010101 01010111

To hide the letter A whose binary value is 01000001, we would replace the LSBs of these pixels to have the following new grayscale values:

1001011**0** 1000110**1** 1101001**0** 0100101**0** 0010011**0**
0100001**0** 0001010**0** 0101011**1**

As from above example, only half the LSBs need to change. The difference between the cover image and the stego image will be hardly noticeable to the human eye.

## 2.1 Mean Squared Error (MSE)

Described as a signal fidelity measure, the goal of a signal fidelity measure is to compare two signals by providing a quantitative score that describes the degree of similarity / fidelity or, conversely, the level of error/distortion between them [5].

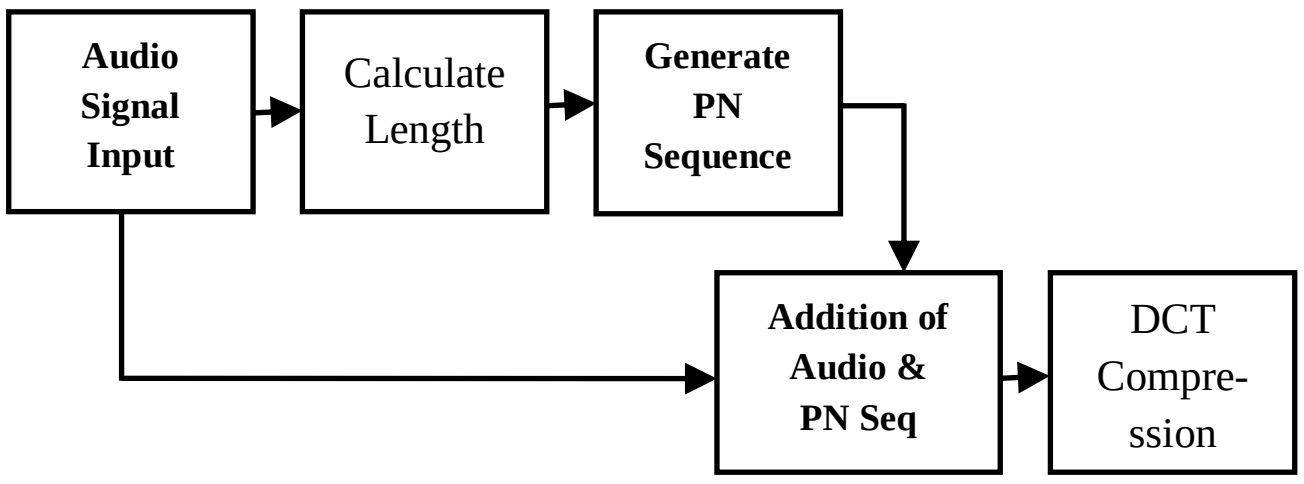

**Fig. 3: General block diagram of embedder**

Suppose that $x = \{x_i \mid i = 1,2,...N\}$ and $y = \{y_i \mid i = 1,2,...N\}$ are two finite-length, discrete signals like images, where N is the number of signal samples (pixels, if the signals are images) and $x_i$ and $y_i$ are the values of the $i^{th}$ samples in $x$ and $y$, respectively. The MSE between the signals is given as

$$MSE(x,y) = \frac{1}{N}\sum_{i=1}^{N}(x_i - y_i)^2 \quad (1)$$

For the steganographic purpose, $x$ is cover image and $y$ is message to be hidden. Hence it will be referred to as error signal, a difference between original image and its watermarked version.

## 2.2 Peak Signal-to-Noise Ratio (PSNR)

For image processing specifically, MSE is converted into PSNR as follows:

$$PSNR = 10\log_{10}\frac{L^2}{MSE} \quad (2)$$

where L is the dynamic range of allowable image pixel intensities, calculated as follows:

$$L = 2^n - 1 \quad (3)$$

Where $n$ is number of allocated bits/pixel. The PSNR is used to evaluate the quality of stego image. For an M x N grayscale image,

$$PSNR = 10\log_{10}\frac{255\times 255\times M\times N}{\sum_{i=1}^{M}\sum_{j=1}^{N}(p_{i,j} - q_{i,j})^2} \text{ dB} \quad (4)$$

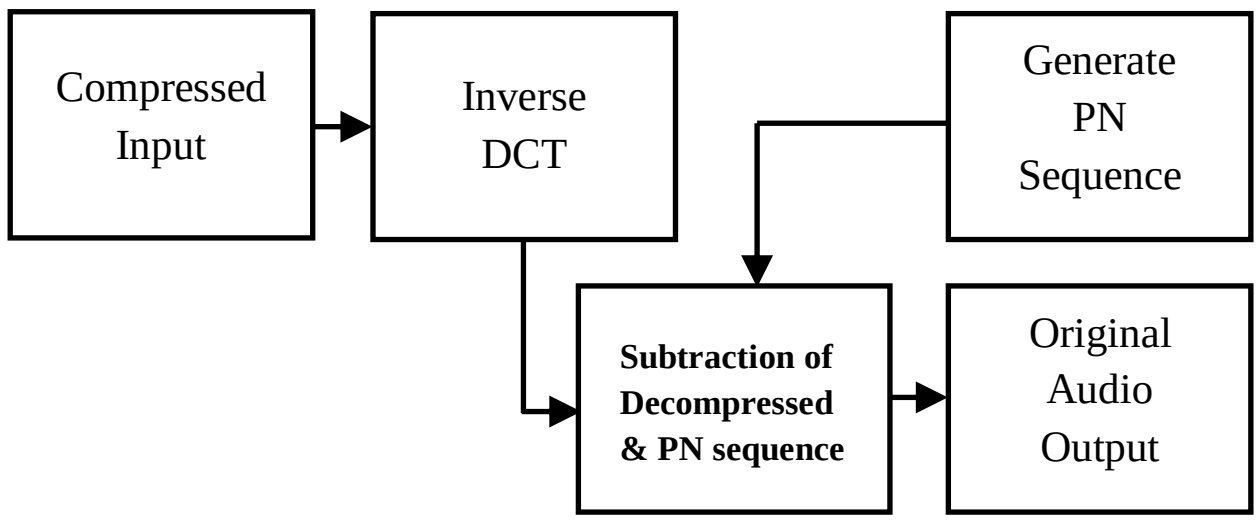

**Fig.4: General block diagram of extractor**

where $p_{i,j}$ and $q_{i,j}$ denote the pixel values in row $i$ and column $j$ of the cover image and the stego image, respectively.

## 2.3 Audio in Audio Watermarking

In this method, both cover and message signals are audio signals. For watermarking purpose, Discrete Cosine Transform (DCT) can be used [6]-[8]. For both audio signals, 1D DCT is sufficient for efficient watermarking. Upon application, the high frequency DCT coefficients of the cover audio file are replaced with the low frequency DCT coefficients of the watermark audio file.

1D DCT is defined as follows:

$$F(u) = \sqrt{\frac{2}{N}}\sum_{i=0}^{N-1}A(i)*\cos\left(\frac{u(2i+1)\pi}{2N}\right)*f(i) \quad (5)$$

Where

$$A(i) = \begin{cases}\frac{1}{\sqrt{2}} & \text{for } u = 0\\ 1 & \end{cases}$$

Otherwise.

$f(i)$ is the input sequence.

## 2.4 Image in Audio Watermarking

In this method, the cover file is audio whereas the message embedded is image signal. It can be performed by using Discrete Wavelet Transform (DWT) on audio signal to be watermarked.

One directional DWT of any stage can be calculated from the coefficients of the DWT of the previous stage using following iterative equations [9]:

$$WL(n,j) = \sum_{m} WL(m, j-1)h_0(m-2n) \quad (6)$$

$$WH(n, j) = \sum_m WL(m, j-1) h_1(m-2n) \qquad (7)$$

where $WL(n, j)$ is the $n^{th}$ scaling coefficient at the $j^{th}$ stage, $WL(n, j)$ is the $n^{th}$ wavelet coefficient at the $j^{th}$ stage, and $h_0(n)$ and $h_1(n)$ are he dilation coefficients corresponding to the scaling and wavelet functions, respectively.

### 2.5 Audio in Image Watermarking

In this type, cover file is image which gets embedded with audio file as message or watermark. It can also be performed by DWT. By calculating DWT coefficients of message files, the coefficients of cover signal are interchanged. As it is similar to method for image in audio watermarking, DWT can be performed by considering equations (6) and (7).

## 3. IMPLEMENTATION STEPS

Two images, cover and message were employed for the purpose. Figures 3 and 4 show the general block diagram of embedder and extractor.For an embedder, the cover image and message image undergo various processes. Both the images should be of same size. Then LSB encoding is performed on them. For experimental purpose, the number of LSB bits to be substituted is determined by user. Let that number is $n$. The bits of message image are shifted by $n$. For cover image, $2^{n-1}$ bits are complemented. Now cover image and message image are added bit-by-bit. This gives the stego image. MSE and PSNR are calculated for comparison purpose.

For an extractor, the stego image is shifted by 8-$n$ bits and then complemented bit-by-bit. This gives the message image.

## 4. EXPERIMENTAL RESULTS

The 2-D (645 x 645, 8 b/pixel) image shown in figure 5(a) was used as cover image and image shown in figure 5(b) was used as message image in order to illustrate the stages of image steganography algorithm and visually assess the quality of the results. Figures 6(a) and (b) are grayscale versions of them respectively.

Fig. 7(a) shows the stego image for $n$ =4. Fig. 7(b) shows the extracted message (for $n$ =4) .

Histograms of cover image and stego image were plotted in fig. 8(a) and 8(b).

For different values of $n$, the values of MSE and PSNR were calculated. Table 1 shows corresponding values of MSE and PSNR for $n$ and time required for each steganographic operation.

When values for MSE and PSNR were plotted for each value of $n$, graphs were obtained as shown in fig. 9(a) and 9(b).

Image in Image

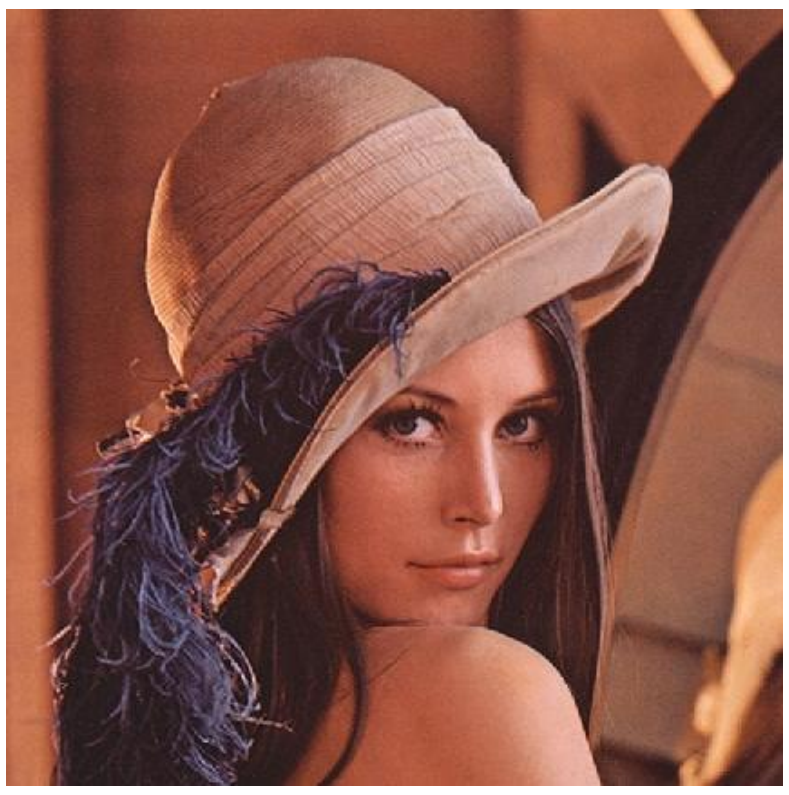

**Fig. 5(a): Cover image (colored)**

**Fig. 5(b): Message image**

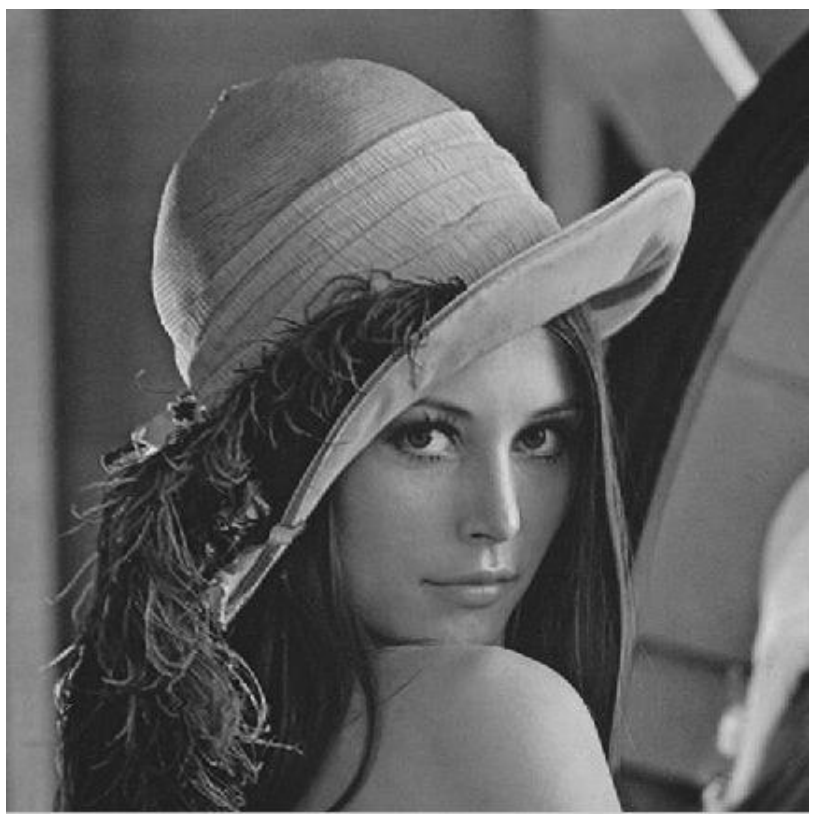

Fig. 6(a): Cover image (grayscale)

Fig.6(b): Message image (grayscale)

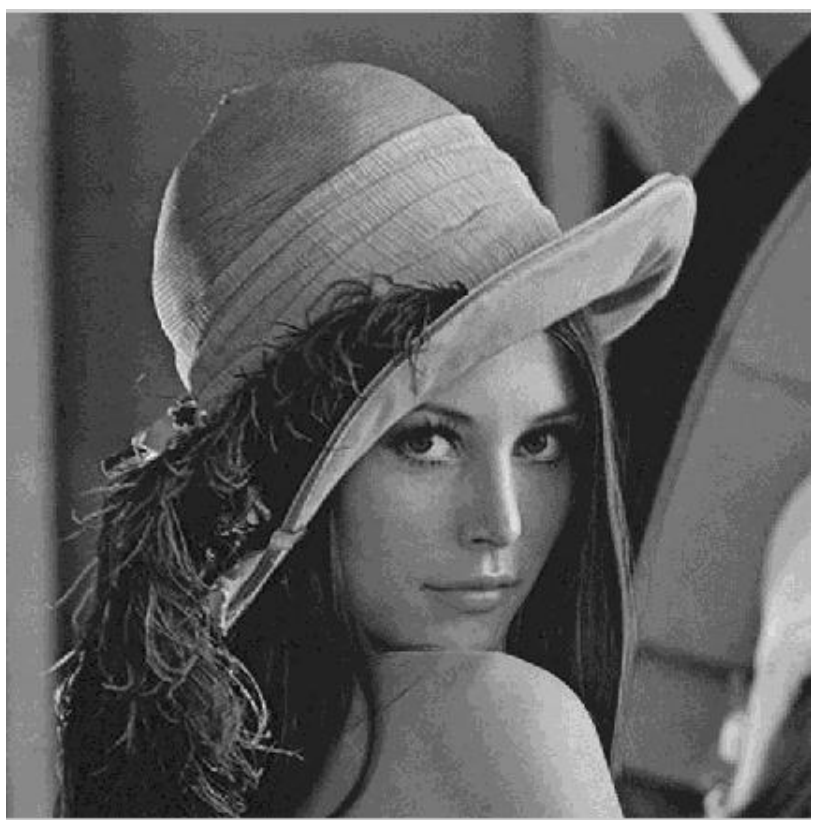

Fig.7(a): Stego image (for $n$ =4)

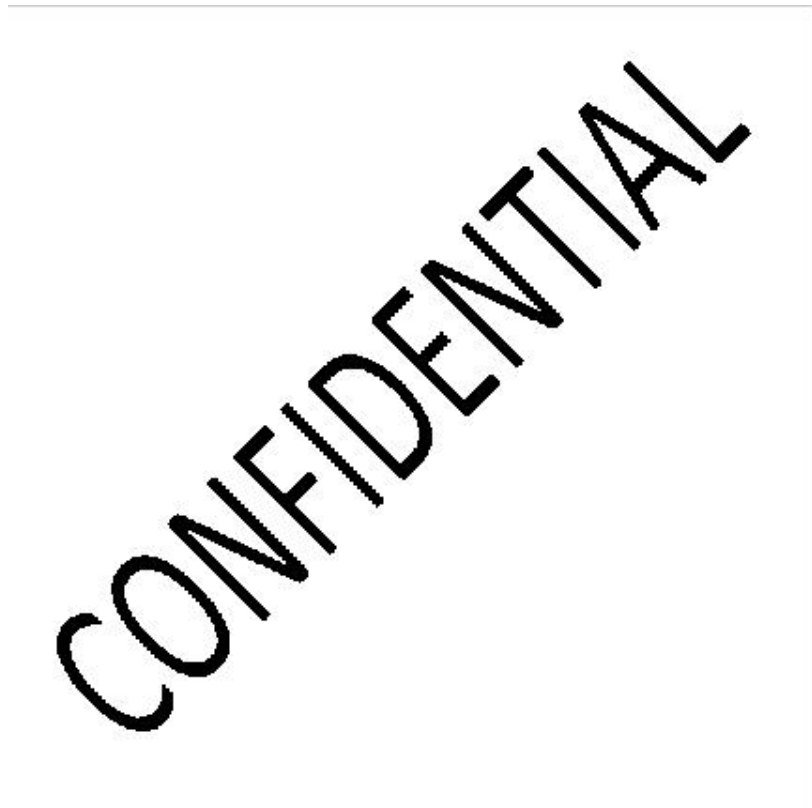

Fig.7(b): Extracted message (for $n$ =4)

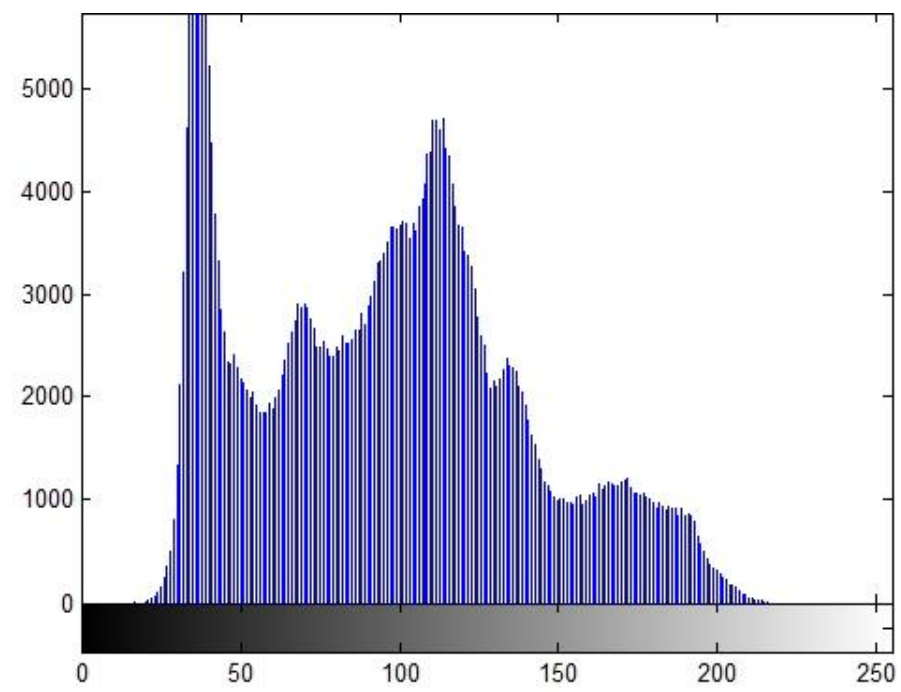

Fig. 8(a): Histogram of cover image

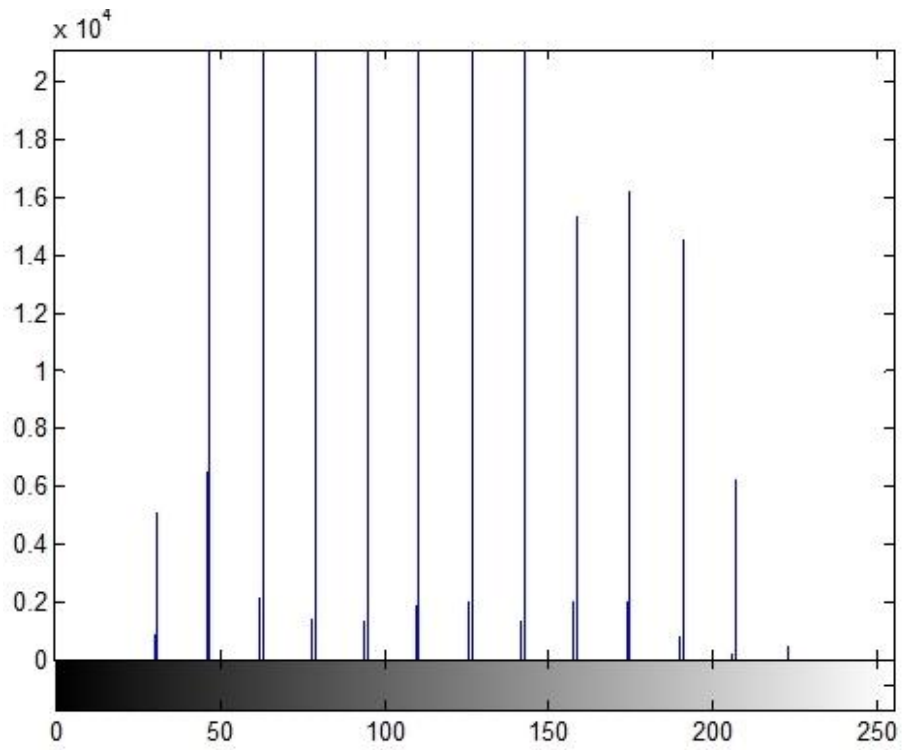

Fig. 8(b): Histogram of stego image

**Table 1: Values of MSE and PSNR along with time taken for each value of $n$**

| n (LSB Bits) | MSE | PSNR (dB) |
|---|---|---|
| 1 | 1.5894e+004 | 14.1790 |
| 2 | 3.8599e+003 | 20.3254 |
| 3 | 915.1061 | 26.5765 |
| 4 | 222.9934 | 32.7083 |
| 5 | 48.5570 | 39.3287 |
| 6 | 8.9208 | 46.6871 |
| 7 | 0.9932 | 56.2210 |
| 8 | 0 | 99 |

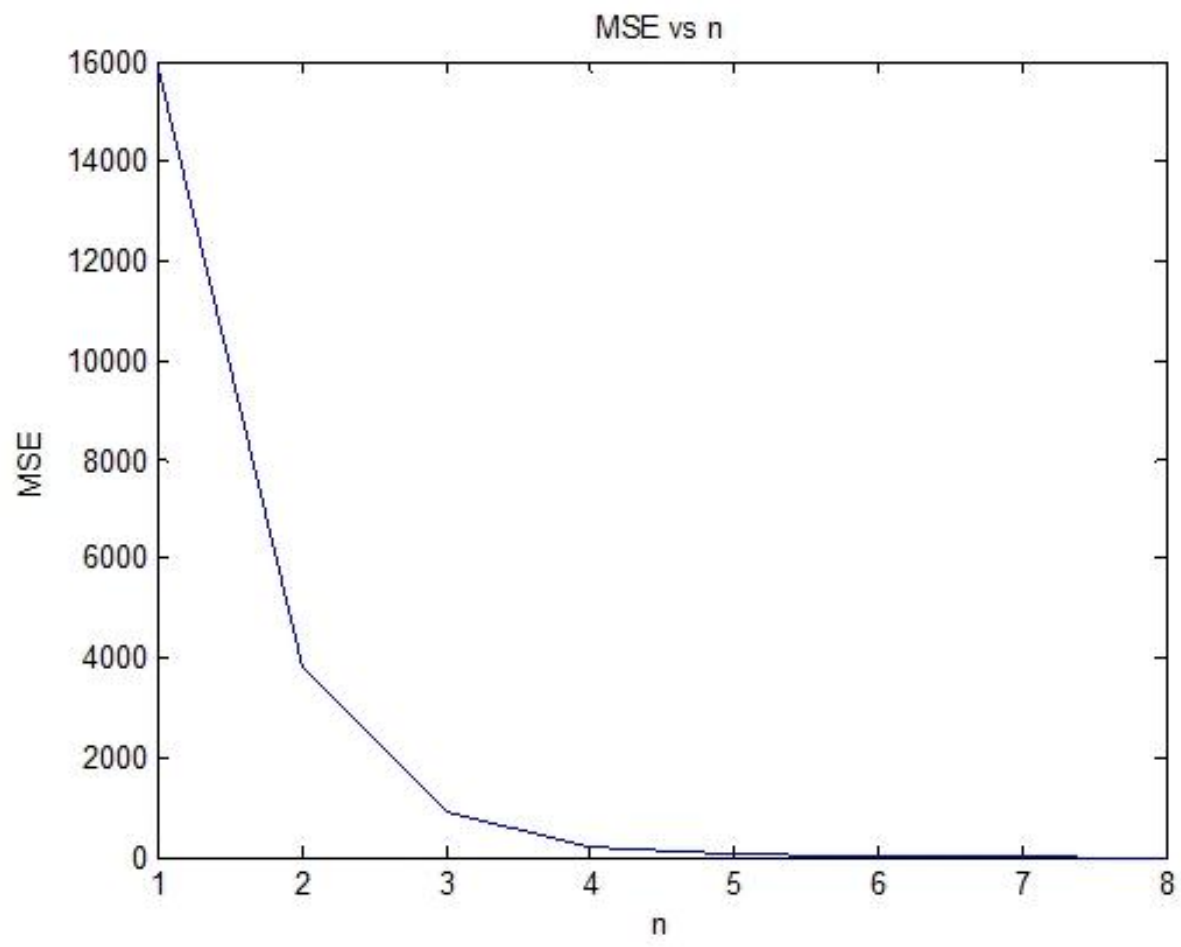

**Fig. 9(a): Plot of MSE vs $n$**

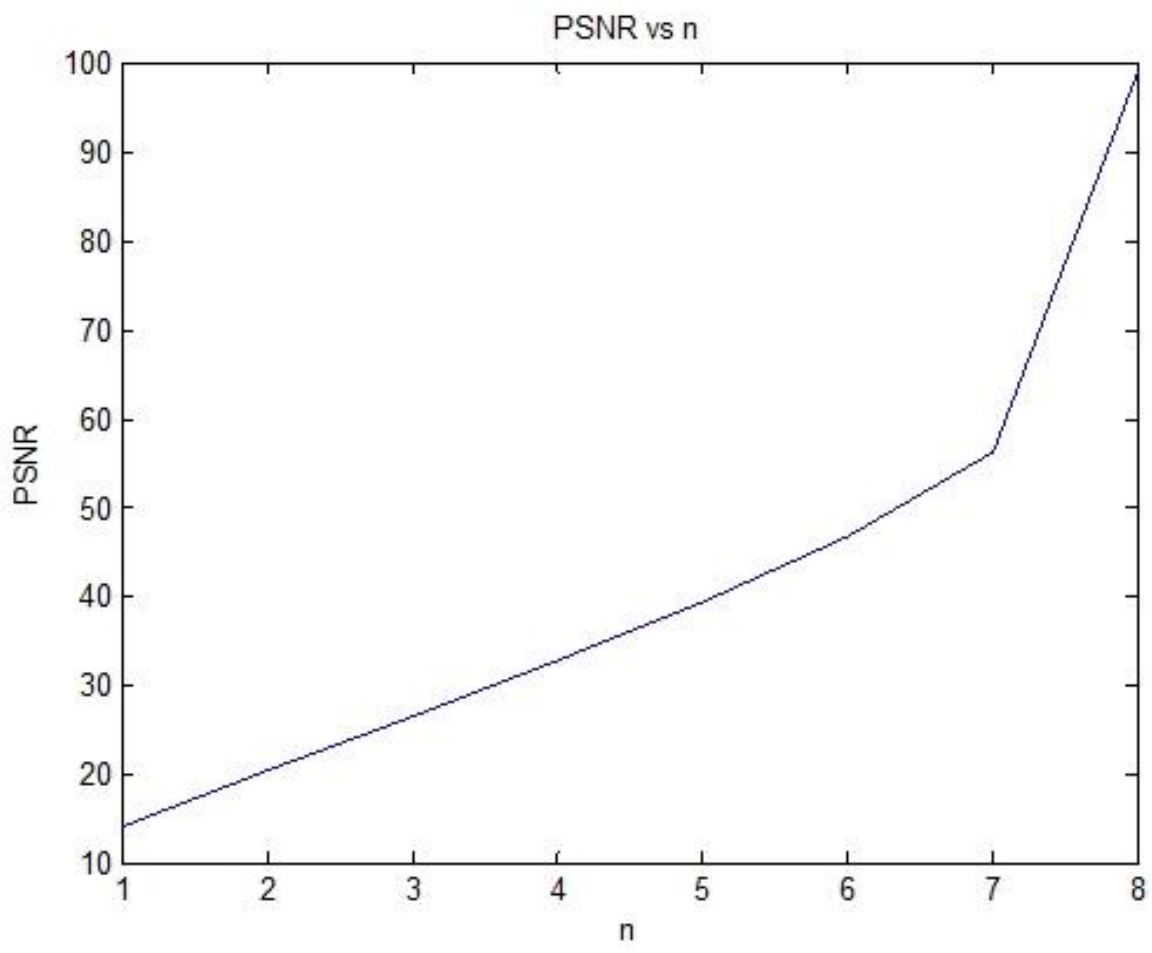

**Fig. 9(b): Plot of PSNR vs $n$**

## Audio in Audio

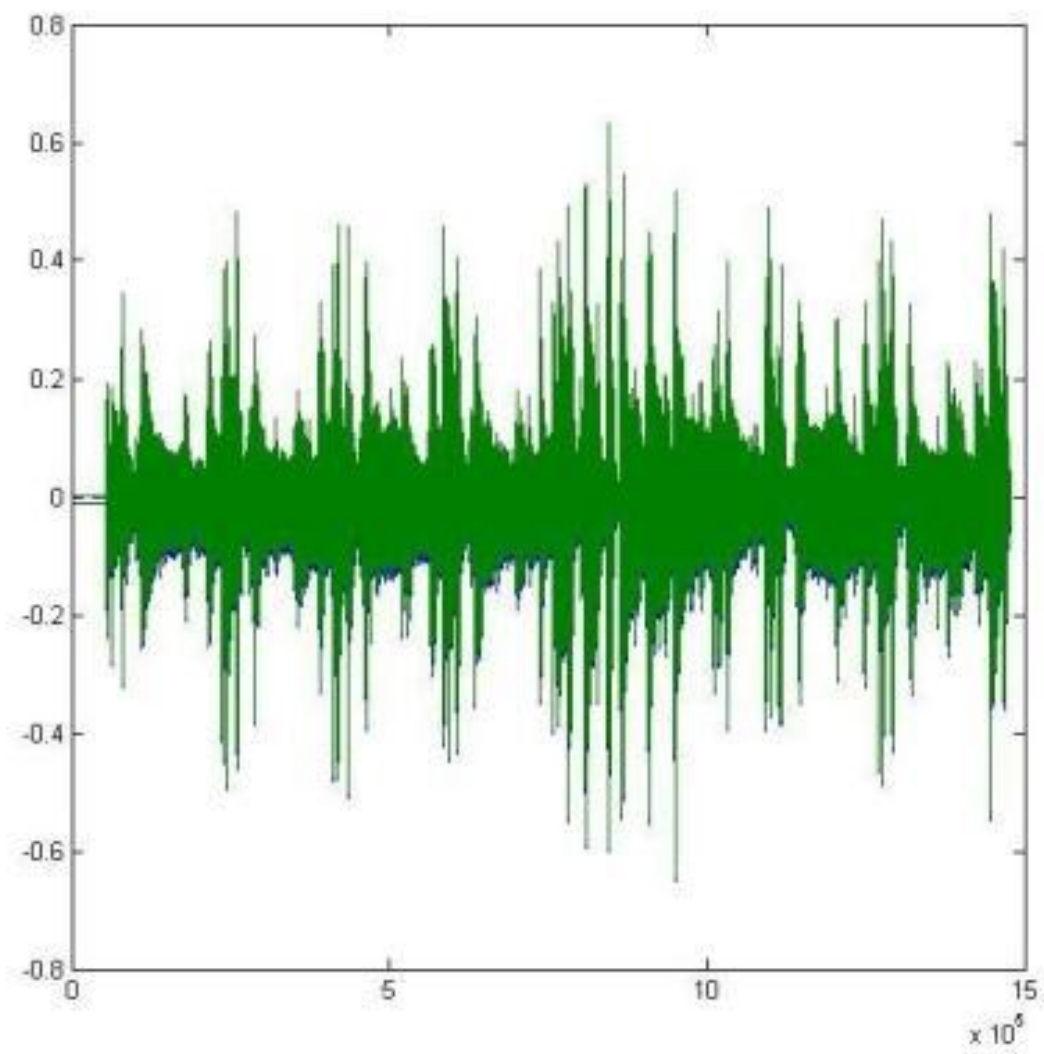

**Fig. 10(a): Original audio signal as cover**

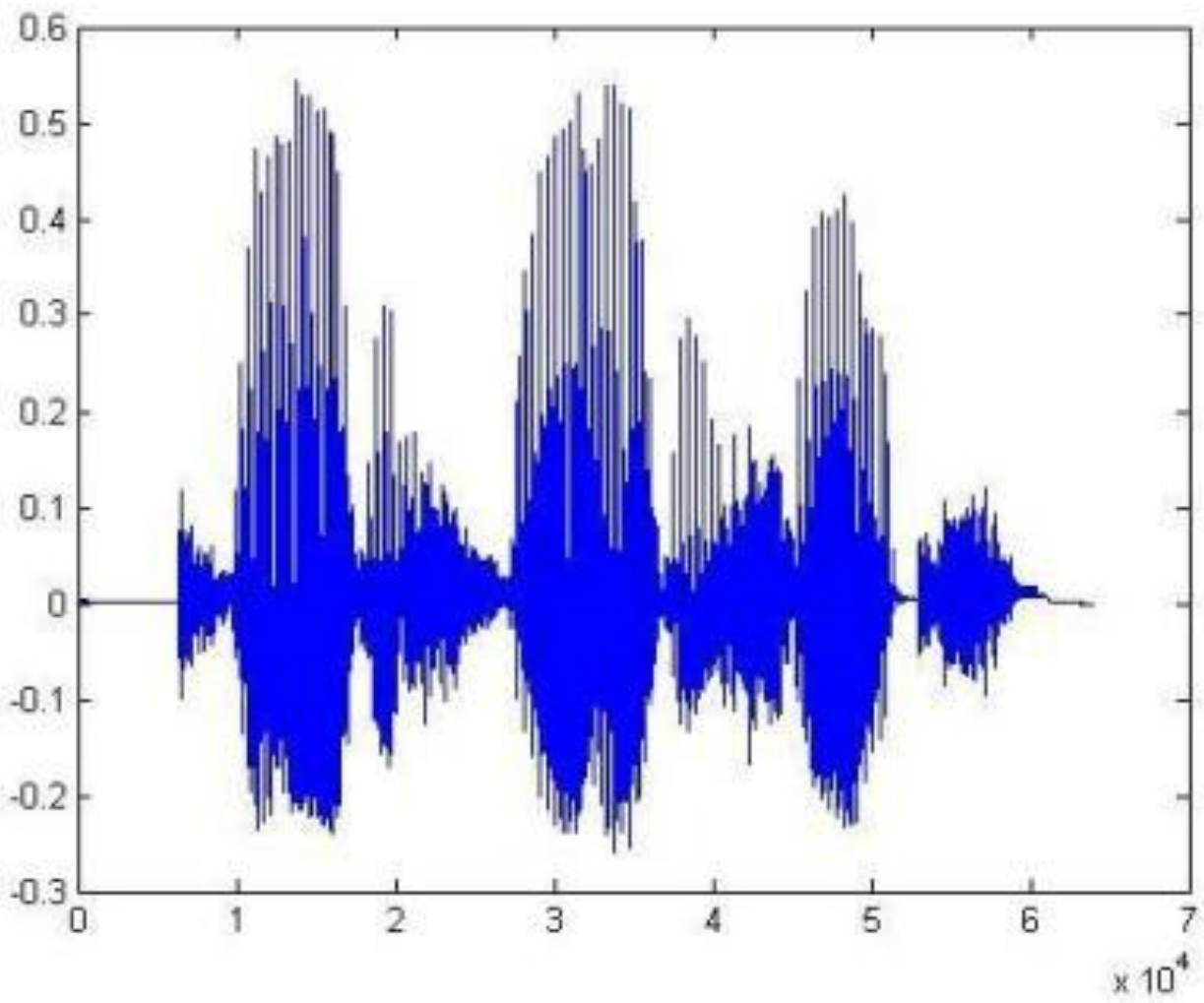

**Fig. 10(b): Watermark audio signal as message**

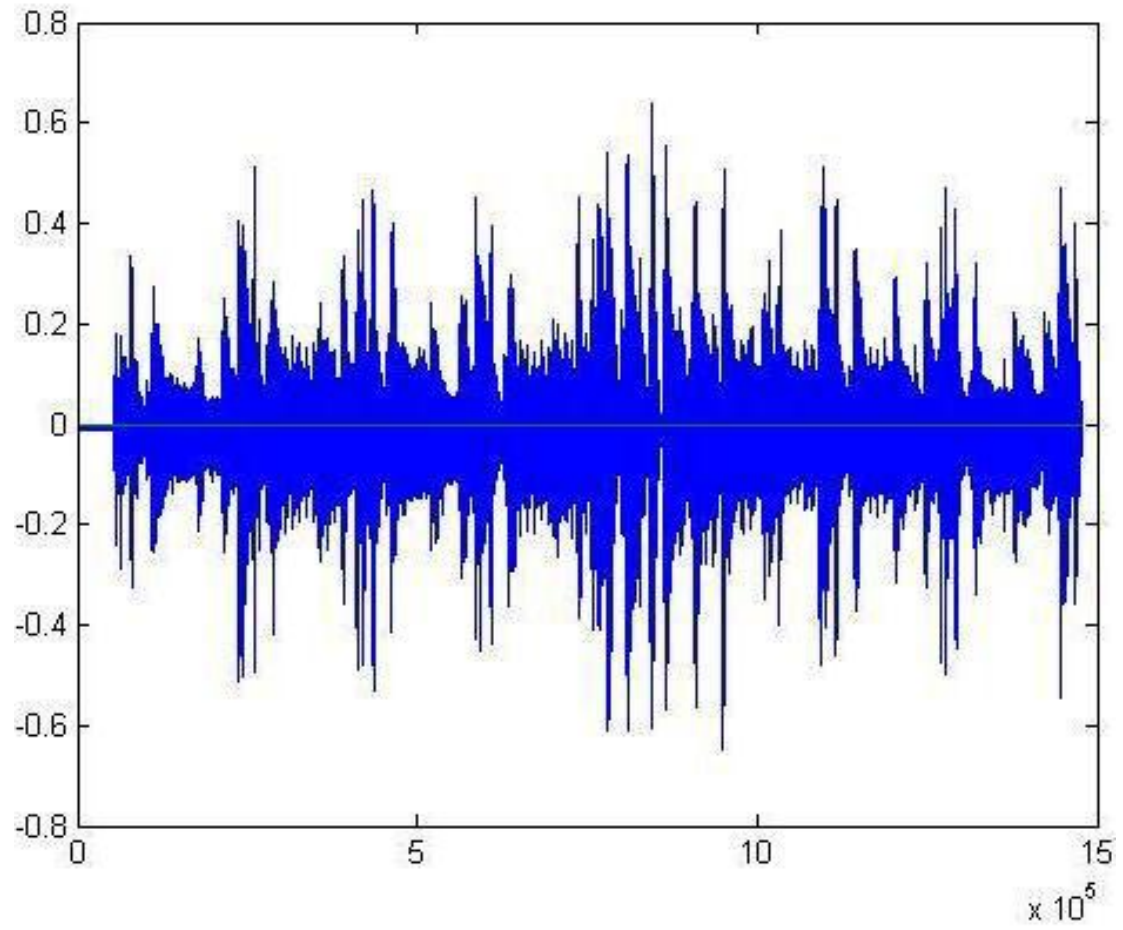

**Fig. 10(c): Audio in audio watermarked signal**

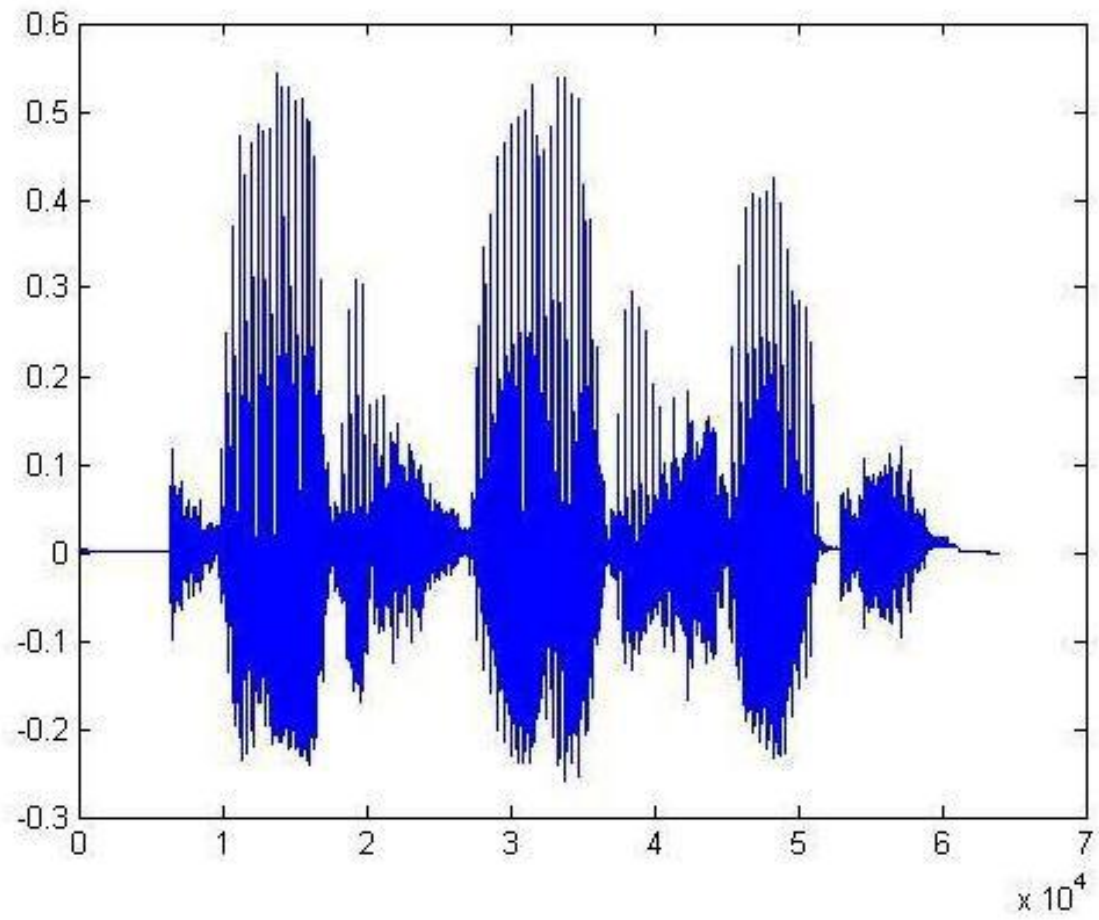

**Fig. 10(d): Recovered audio watermark signal**

Fig. 10(a) shows the original audio signal used as cover, 10(b) shows watermark audio signal used as message. Fig. 10(c) shows audio in audio watermarked signal and fig. 10(d) shows recovered audio watermark signal.

## Audio in Image

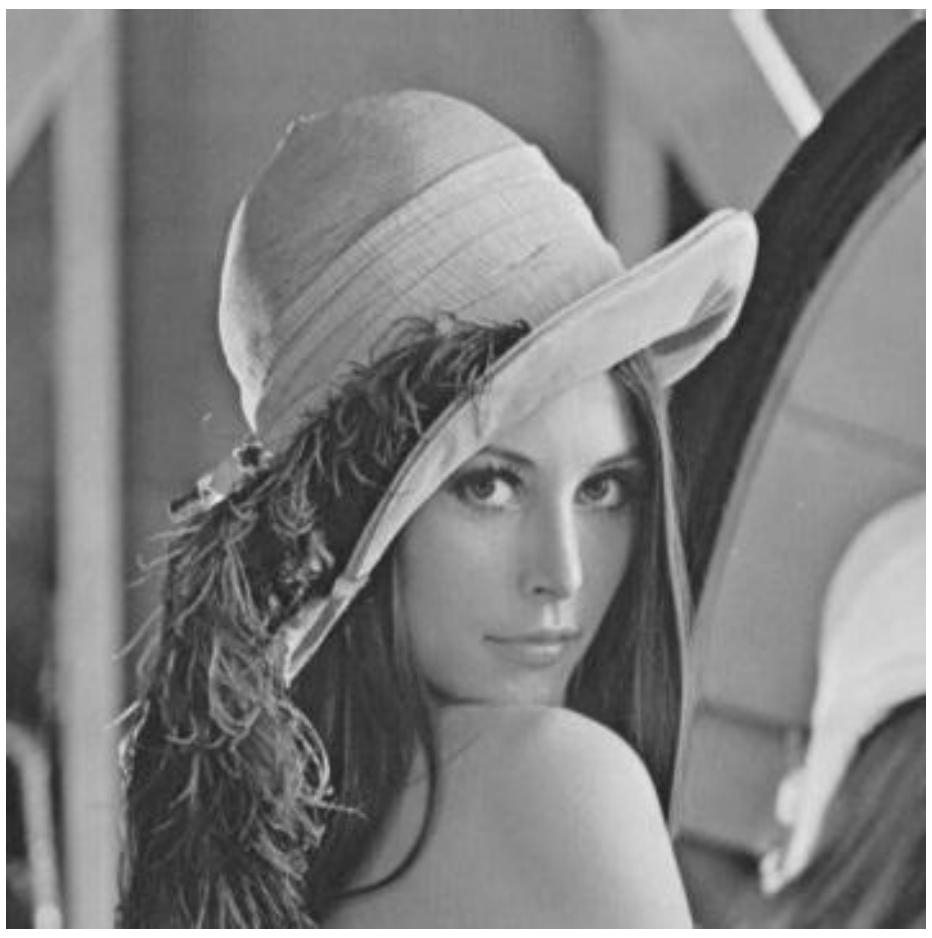

**Fig. 11(a): Original image signal as cover**

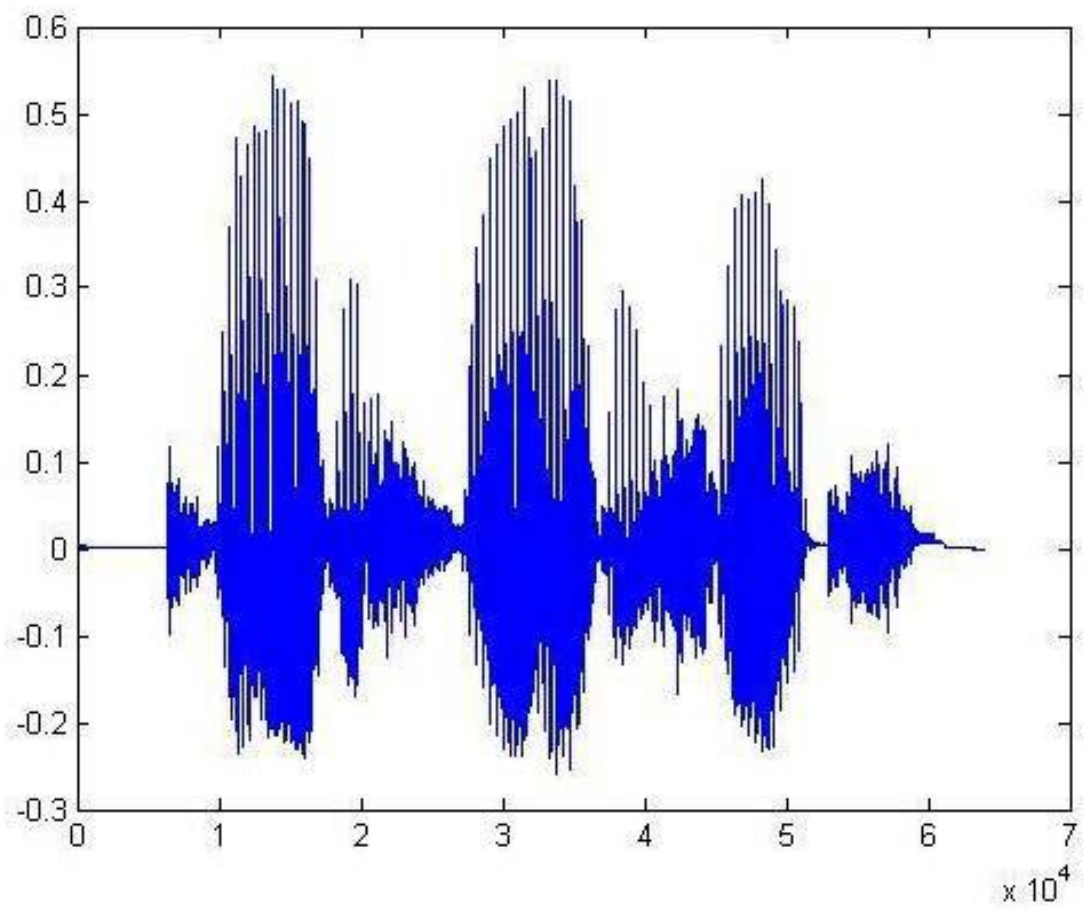

**Fig. 11(b): Watermark audio signal as message**

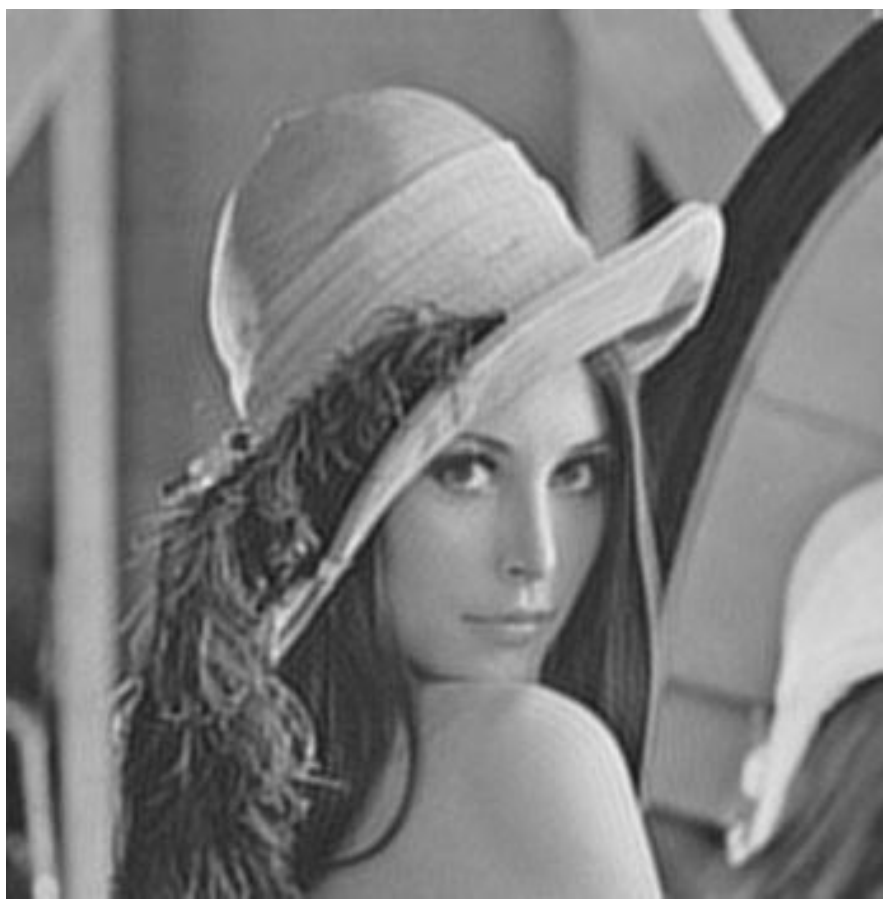

**Fig. 11(c): Audio in image watermarked signal**

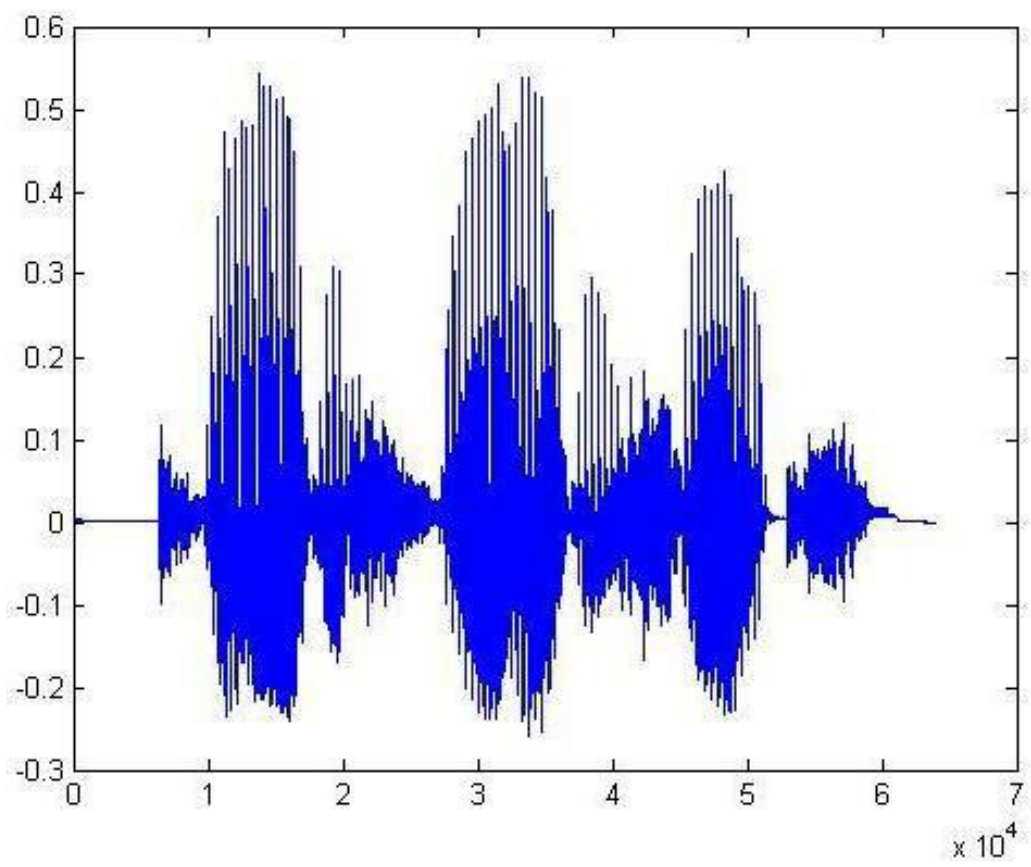

**Fig. 11(d): Recovered audio watermark signal**

Fig. 11a) shows the original image signal used as cover, 11(b) shows watermark audio signal used as message. Fig. 11(c) shows audio in image watermarked signal and fig. 11(d) shows recovered audio watermark signal.

Image in Audio

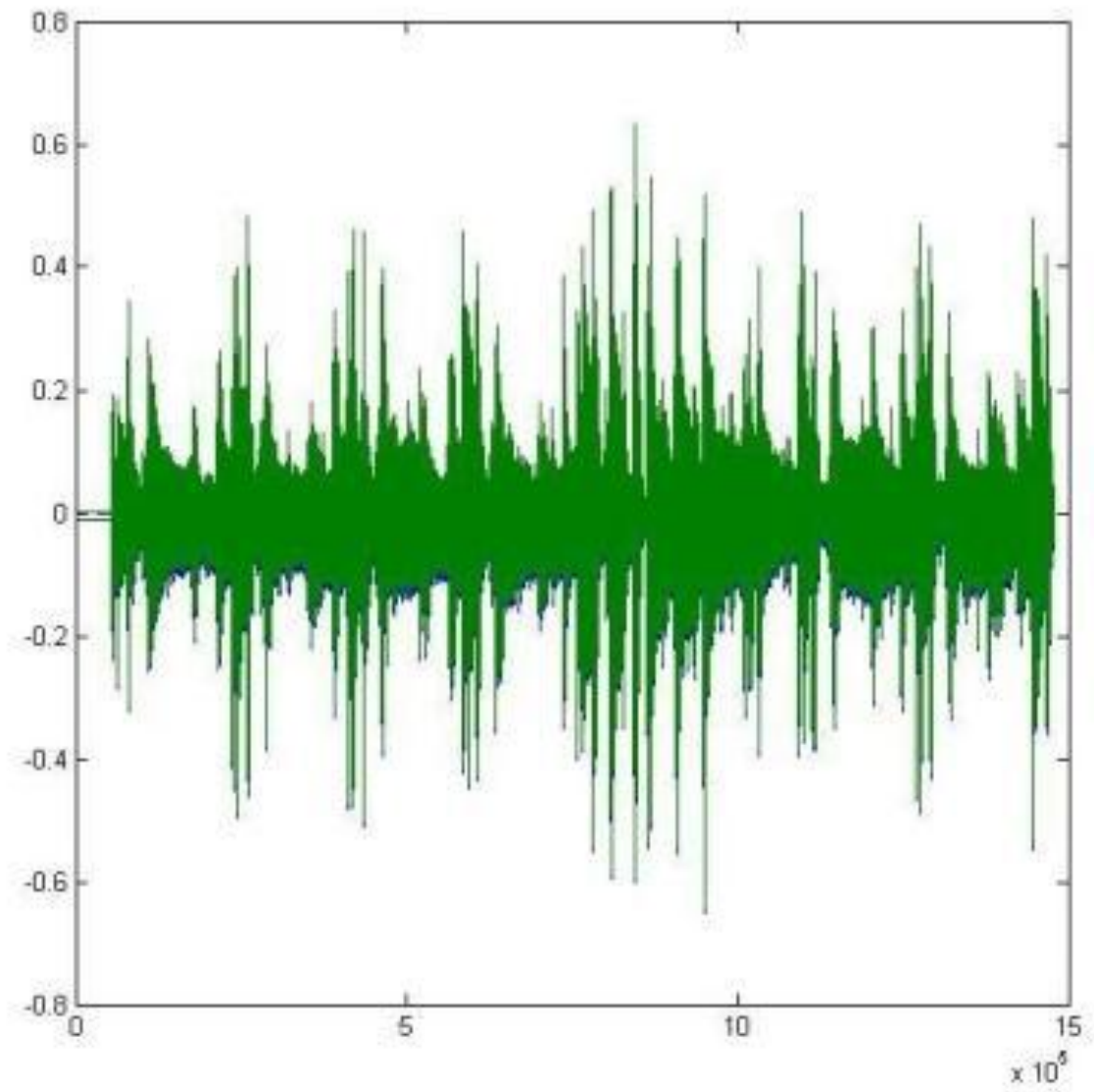

**Fig. 12(a): Original audio signal as cover**

**Fig. 12(b): Watermark image signal as message**

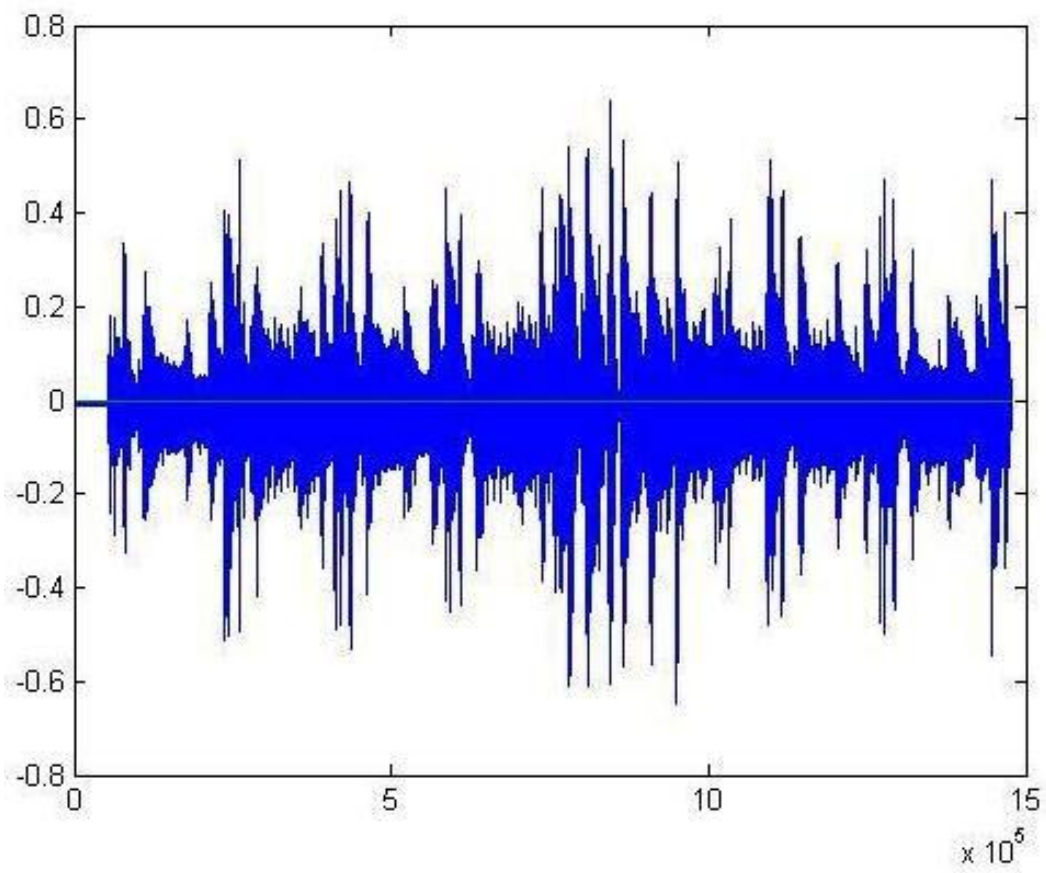

**Fig. 12(c): Image in audio watermarked signal**

**Fig. 12(d): Recovered image watermark signal**

Fig. 12(a) shows the original audio signal used as cover, 12(b) shows watermark image signal used as message. Fig. 12(c) shows image in audio watermarked signal and fig. 12(d) shows recovered image watermark signal.

## 5. CONCLUSION

The main advantage of the method is a very high watermark channel capacity; the use of only one LSB of the host audio sample gives capacity of 44.1 kbps. The obvious disadvantage is the extremely low robustness of the method, due to fact that random changes of the LSBs destroy the coded watermark. In addition, it is very unlikely that embedded watermark would survive digital to analogue and subsequent analogue to digital conversion. Since no computationally demanding transformation of the host signal in the basic version of this method needs to be done, this algorithm has a very small algorithmic delay. This permits the use on this LSB in real-time applications. This algorithm is a good basis for steganographic applications for audio signals and a base for steganalysis.